# Hidden in snow, revealed in thaw
## Cold boot attacks revisited

Jos Wetzels

`a.l.g.m.wetzels@student.tue.nl`

**Abstract.** In this paper, we will provide an overview of the current state-of-the-art with regards to so-called cold boot attacks, their practical applicability and feasibility, potential counter-measures and their effectiveness.

**Keywords:** Cold boot attack, Side-channel attack, Computer forensics, Memory acquisition

## 1 Introduction

Side-channel attacks [1], which focus on physical implementations of cryptosystems and the potential flaws arising from them rather than cryptographic weaknesses, are a well-known study subject within the field of computer security. One such side-channel attack is the so-called cold boot attack (CBA) [2], which first came to widespread public attention in 2008 [3]. In short, cold boot attacks rely on the fact that Random Access Memory (RAM) exhibits so-called data remanence[4]. This means that, even though both Dynamic and Static RAM are volatile (in the sense that data will be lost eventually upon power removal) memory types, residual digital information stored on such memory is present for some time even after power has been removed. This property enables an attacker to potentially recover information (such as cryptographic keys, passwords, etc.) that compromises system security. A CBA exploits this property by initiating a so-called cold boot, where power to the target machine is cycled off and on again (without allowing the operating system to properly exit and thus potentially erase or overwrite sensitive memory contents) after insertion of a bootable disk containing specialized software to immediately dump raw pre-boot memory contents. Subsequent analysis against the dumped memory is then performed to recover data of interest to the attacker. With certain types of memory, the attack time window can be extended by cooling them. Of course, it goes without saying that execution of a CBA requires physical access to the target machine on part of the attacker.

## 2    Cold Boot Attacks in-depth

An especially interesting target for CBAs are so-called Full-Disk Encryption (FDE) systems [5]. FDE systems protect sensitive information on storage media by fully encrypting all information present on them, sometimes including the bootable operating system partitions. Benefits of such an approach include the fact that everything on a storage medium, including swap space and temporary files (which potentially contain sensitive data), is encrypted as opposed to only those files considered sensitive by the user. In addition, FDE offers the user the option of immediate effective data destruction (though secure wiping or physical destruction is advised when considering potential future attacks) through the destruction of the cryptographic keys used to encrypt the storage medium.

As society continues to increasingly rely on information technology, the amount of digitized sensitive information as well as the potential impact of its loss increases accordingly. Major cases of loss and theft of storage media containing sensitive information have spurred on the adoption of FDE systems [6] such as Truecrypt or BitLocker [7,8]. As such, attackers seeking to obtain access to sensitive information (for espionage, criminal or law enforcement purposes) face an increasingly difficult task.

FDE systems, however, have the tendency to store sensitive information (such as cryptographic keys or information from which these can be derived) in-memory during runtime for user convenience and speed purposes. As such, an attacker with physical access to a running (but potentially locked) or suspended FDE protected system could perform a CBA to extract this sensitive information from memory and hence bypass the protection offered by FDE systems. Such a scenario poses a particular threat to users of mobile systems (such as laptops, netbooks, smartphones, etc.) left unattended in a locked or suspended state or those who are targeted by law enforcement raids while their systems are running. Needless to say, not only FDE systems are targeted with CBAs as target systems potentially hold more sensitive information in-memory, such as (online account) passwords or cryptographic keys to other applications.

The basic CBA scenario is roughly as follows:

1. Take necessary steps (such as cooling the memory and taking countermeasures into account) to get a suitable attack time window and attack conditions
2. Obtain an image of the residual memory contents
3. Identify and reconstruct sensitive information from the obtained memory image

We will discuss these steps in the following sections, beginning with the data remnance phenomenon underlying CBAs.

## 2.1 Decay and remnance

Given the intended targets of CBAs, we will mostly concern ourselves with DRAM, which is a type of memory [9] where each bit of data is stored in a separate capacitor that is either charged or discharged (representing the two binary values each cell can take). As time passes, the capacitors will slowly discharge and decay to their ground state (which is either zero or one, depending on the capacitor's wiring). Because of this decay, cells must be periodically refreshed (ie. memory contents are read and re-written so cells hold their value) according to a certain refresh rate (which is usually within the order of milliseconds).

However, as mentioned in Section 1, contrary to (what used to be) popular belief, DRAMs do not immediately lose their state if they aren't regularly refreshed. In fact, most DRAM modules retain their state largely in-tact (without refresh, without power and even when removed from a motherboard) for periods of up to thousands of refresh intervals [10]. Halderman et al.[3] observed that at normal operating temperatures (varying from 25.5°C to 44.1°C) DRAM saw a generally low rate of bit corruption for a period of several seconds, followed by a period of very rapid decay, with newer memory technologies tending to decay faster. In general, despite the fact that these decay patterns varied between models, they are highly non-uniform and both the ground states to and the order in which various different cells decayed almost invariably proved very predictable.

It has been widely documented that memory operating at lower temperatures displays significantly higher retention times [11,12], thus aiding attackers in increasing their attack time window. Using simple cooling techniques such as spraying from an upside-down container of canned air (so that it dispenses its contents in liquid form and not as a gas), an attacker can obtain surface temperatures of -50°C. This ensures that less than 1% of the memory contents are decayed, even after a period of 10 minutes without power. Subsequent tests showed that submerging DRAM modules in liquid nitrogen to obtain surface temperatures of ~ -196°C meant a decay of only 0.17% of the memory contents after a period of 60 minutes without power, suggesting removed DRAM modules stored with proper cooling may be susceptible to reliable recovery after periods of multiple hours or even days.

In addition to decay, attackers seeking to perform a CBA are faced with the fact that upon booting the system the BIOS often overwrites parts of the memory, though this is usually only a small fraction (in the order of a couple of megabytes) of the memory often located at the lowest part of the address space. Another problem arises from the fact that some system BIOS perform a so-called Power-On Self Test (POST) [13] which overwrites large sections (or sometimes all) of the memory. Fortunately for an attacker, this test can usually be bypassed by setting certain BIOS options. A final problem comes in the form of systems supporting so-called Error-Checking & Correction (ECC) memory where the BIOS sets memory to a known state (thus effectively erasing its previous contents) to avoid errors from occurring upon reading uninitialized data, a feature that cannot be overridden. Attackers seeking to perform a CBA on such a system would have to transfer the ECC memory modules in question to a non-ECC system, potentially benefiting from ECC's increased error-resistance.

## 2.2  Image acquisition

In order to obtain an image of residual memory contents an attacker doesn't need any special equipment. One problem, however, is that upon system boot, the memory controller initiates DRAM refreshing and hence reads and rewrites each bit, fixing their values and halting decay. Since (as discussed in section 2.1) booting a system will invariably overwrite some parts of memory, an attacker should seek to limit this by booting from a small, special-purpose operating system which dumps memory contents to an external storage medium after (either warm or cold) boot. The imaging tools presented by Halderman et al. [3] (booting from PXE, USB Drive, EFI or iPod) provide such functionality, using only a small amount of RAM and being located at memory offsets chosen so that data structures of potential interest remain unaffected and saving the memory dump to the medium from which they were booted.

Depending on counter-measures put in place (see section 4) and the attacker's access to the target system, the aforementioned imaging tools could be utilized in either of the following scenarios [3], presented in order of ascending sophistication and resistance to counter-measures:

1. **Warm boot:** This scenario involves a simple warm boot, using the OS' restart functionality which, though preventing decay, gives any running software the opportunity to perform a memory wipe before shutting down.

2. **Cold boot:** This scenario (the CBA proper) involves briefly interrupting and restoring power or using the system restart switch so that only little to no decay (depending on the particular memory's properties) takes place while avoiding giving any running software the opportunity to perform a memory wipe.

3. **Transferring memory modules:** Even if the previous two scenario's (such as due to counter-measures erasing memory contents upon boot) are infeasible, an attacker could still physically replace the memory modules of the target system and dump an image of their contents using another, attacker-controlled, system. Though such an approach exposes memory modules to greater decay than the previous scenarios, section 2.1 shows that an attacker could cool down the memory module before shutting off power in order to extend their attack time window (even up to several minutes) and limit decay. One additional benefit of this approach is that an attacker could place a removed primary memory module as a secondary memory module on their target machine and thus map the contents to a different section of the address space, preserving memory contents otherwise overwritten during booting.

### 2.3 Key reconstruction

Despite the fact that memory contents can be recovered using a CBA with relatively few errors even a small amount of errors can complicate the extraction of cryptographic keys significantly. As such, research into CBAs has included optimizing methods for correcting errors in cryptographic keys for both symmetric and asymmetric algorithms (we will use the AES and RSA cases as illustrative examples). The most naïve approach (which quickly becomes infeasible even under conditions of minor decay) would be a bruteforce search over the set of keys with low hamming distance from the extracted (decayed) key. More sophisticated reconstruction algorithms use the fact that most encryption solutions (see section 3) store either pre-computed key schedules [14] or extended forms of the key in-memory. A key schedule is a set of keys derived from the master key, each of which is used in the separate cipher rounds. In order to enhance performance (as part of an efficiency-security trade-off), key schedules are often pre-computed and stored in memory (at least as long as the encrypted volume is mounted). Since the round keys are derived from the master key through the cipher's scheduling operations, the key schedule can be seen as an error-correcting code [15] and the key reconstruction problem is essentially a coding-theoretical decoding problem. Using such an approach, keys can be reconstructed far more efficiently under conditions of far worse decay. An additional benefit of such methods is that keys can be recovered without having to test decrypted ciphertext against the resulting plaintext, using data derived from the key as an indication of the likelihood of its correctness.

It is noted [3] that in most cases memory bits tend to decay to a predictable ground state, with only a minority flipping in the reverse direction. Under the assumption that an attacker has no knowledge of decay patterns except for the ground state, decay can be modelled as a binary asymmetric channel for various levels of decay described by probability $\delta_0$ (the chance of a 1 flipping to a 0) and some fixed probability $\delta_1$ (the chance of a 0 flipping to a 1). As time passes and decay progresses, $\delta_0$ will approach 1 while $\delta_1$ remains fairly constant (usually under 0.1%). Ground state decay probability $\delta_0$ can be approximated from the acquired memory data through the fraction of 1s and 0s (since cryptographic key data is assumed to be uniformly distributed). This yields two possible scenarios, that of *perfect* asymmetric decay (where $\delta_1 = 0$) and that of *imperfect* asymmetric decay (where $\delta_1 > 0$, usually between 0.05% and 0.1%). Under conditions of perfect asymmetric decay all 1s present in the recovered key schedule can be presumed correct with certainty, while this does not hold for those under conditions of imperfect asymmetric decay. Of the various key reconstruction approaches some can only handle conditions of perfect asymmetric decay, while others can handle both scenarios (though with significant performance penalties in the latter case).

#### 2.3.1 AES

The Rijndael cipher, selected as the Advanced Encryption Standard (AES) [18] in 2001, is present in a wide range of cryptographic solutions (often as the default op-tion). AES (128-bit) key expansion consists of 11 round keys of 4 words (128-bits) each, the first round key being the master key itself. Every remaining key schedule word is the product of either a) a XOR operation between two key schedule words or b) the key schedule core operation performed on a key schedule word with the result XORed with another key schedule word. Over time, various approaches to AES key reconstruction have been suggested:

1. *Algorithmic*: Instead of trying to correct entire keys at once, the AES key reconstruction method proposed by [3] works on smaller sets of bytes one at a time, something that is possible because of the linearity of the AES key schedule. This way an optimal decoding can be bruteforced for small enough sequences that are still large enough for the decodings to be useful to overall key reconstruction. After obtaining a list (ordered by likelihood) of possible decodings for those smaller key segments these can be combined into the full key to be checked against the key schedule. The running time of the algorithm is limited by the number of combinations to be checked and while this is still more or less exponential in the number of errors, it is a significant improvement over bruteforcing. It is noted that if $\delta_0$ or $\delta_1$ is sufficiently small it is highly likely a unique solution will be found as well as that the algorithm is capable of reconstructing half of keys with 30% decayed data within 30 seconds.

   Tsow [19] proposed an improved AES key reconstruction method capable of recovering almost all keys with 70% decayed data in less than 20 minutes. The increased speed of the method allows an attacker to enumerate all keys satisfying decay and key schedule constraints (and hence determine whether a found solution is unique or not) within reasonable time (rather than halt on the first match) for decay rates up to 60%. The method proposed by Tsow, which is a recursive branch-and-bound [20] depth-first tree search, works by building up an (initially empty) recovered key schedule by guessing an additional byte $b$ in every recursive call, subsequently computing all bytes that can be derived from the bytes established so far. If this recovered key schedule still satisfies the error model constraints, the recursive functions descends further down this tree-branch. Otherwise (or upon not finding a valid key schedule after the full branch has been investigated) the next value for $b$ is tried. The order in which byte positions are guessed is optimized (using path prioritization with a set of heuristics) to maximize the number of implied schedule bytes and thus minimize the number of compatible candidates at every stage. Riebler et al. [21] expanded upon the work by Tsow by translating the latter's method to a state machine (with optimized datapaths for faster processing of the most commonly accessed tree levels) and implementing it with an FPGA [22], supporting both perfect and imperfect asymmetric decay models. Riebler et al. provided a software implementation outperforming that of Tsow by 3 times and a hardware implementation outperforming Tsow's software implementation by 6 times under perfect asymmetric decay and by 27 times under imperfect asymmetric decay, making AES key reconstruction under conditions of high, imperfect asymmetric decay feasible.

2. *Max-PoSSo*: Further improvements in AES key reconstruction (with applications to Serpent and Twofish as well) have been made through solving polynomial systems with noise based on mixed integer programming [23]. Albrecht et al. [24] note that the key reconstruction situation (the recovery of the original key from a set of decayed round keys) can be cast as a system where every bit of every round key corresponds to a Boolean polynomial equation with the origin key bits as variables with additional noise stemming from decay. The problem of key reconstruction is thus cast as a so-called *Partial Weighted Max-PoSSo (Polynomial system solving)* problem, which is then converted into mixed integer programming problems (since these deal with minimizing or maximizing functions in several variables under constraints of linear (in)equality and partial integrality) solved using the SCIP (Solving Constraint Integer Problems) solver [25]. While it is noted Max-PoSSo problems are analogous to so-called Max-SAT (boolean satisfiability) problems (see below), the former better preserve the algebraic structure of the problem. The method employed by [24] (which can be applied if perfect asymmetry cannot be assured, ie. if $\delta_1 \neq 0$) achieves a maximum success rate of 70% under conditions of 15% decay (taking 11.77 seconds on average) and worst-case non-zero success rates for up to 50% decay (taking 3074.36 seconds on average). Huang et al. [27] presented additional improvements in solving polynomial systems with noise by using a so-called *Incremental Solving and Backtracing Search (ISBS)* method. ISBS tries to incrementally solve a noisy polynomial system in increasing order of hamming weight, using the Characteristic Set (CS) method [27] to improve overall efficiency. The ISBS method achieves a maximum success rate of 75% under conditions of 15% decay (taking 0.07 seconds on average) and worst-case non-zero success rates for up to 50% decay (taking 772.02 seconds on average), with a minimum success rate of 23% compared to the 8% of [24].

3. *SAT-Solving*: Kamal et al. [28] observed that the relations between the AES key schedule round keys can be easily cast as a Boolean Satisfiability (SAT) problem [29], which consists of checking whether there exists an assignment of Boolean values to the variables within a Boolean formula such that the formula yields true. Although SAT-solving is NP-complete, many real life SAT problems can be effectively solved using heuristics, as is done by so-called SAT solvers. SAT solvers have many applications in cryptography [30] and as such Kamal et al. were able to use CryptoMiniSAT [31] to recover 128-bit AES keys under conditions of 70% decay with a worst case of less than 3.5 minutes and average and median recovery times of 1.2 and 0.36 seconds, with promising results for up to 74% decay. Liao et al. [32] presented improvements in using SAT-solving techniques for AES key reconstruction purposes by taking into account the situation of imperfect asymmetric decay (ie. $\delta_1 \neq 0$) explicitly not treated by Kamal et al. This was done by using a Max-SAT solver which considers the key bit relations as hard constraints and the charged state bits as soft constraints (accounting for a $\delta_1 > 0$), attempting to satisfy all hard constraints and maximizing the amount of satisfied soft constraints (as opposed to regular SAT solving which seeks to satisfy all constraints and merely reports the problem as unsatisfiable if this is not possible). In addition, the Max-SAT approach outperformed the regular SAT approach as well, being capable of reconstructing AES keys under conditions of 76% decay (and for $\delta_1 = 0.1\%$) within 300 seconds on average and doing so almost 4 times faster than the regular SAT approach.

### 2.3.2 RSA

RSA [33] is one of the most widely used public-key cryptosystems in the world and features in many cryptographic solutions. An RSA key pair consists of a public key (comprised of modulus N and public exponent e) and a private key (comprised of private exponent d and redundant optional values such as N's prime factors p and q, d mod (p-1), d mod (q-1) and $q^{-1}$ mod p). These optional values are stored in order to increase computation speed. When reconstructing RSA key material from decayed memory an attacker has no key schedule to rely on (as is the case for DES, AES, etc.) but is effectively faced with a partial key exposure scenario, with bits of the secret key and optional values being available with some probability of error. As noted by Halderman et al. [3] there have been various proposed approaches [34,35,36,37] to effectively reconstruct partially exposed RSA private keys, most of which are based on Coppersmith's [35] approach which involves finding solutions to a system of polynomial equations using so-called lattice basis reduction [38] given a known contiguous fraction of (least- or most significant) key bits. Since decayed key material acquired with a CBA might have errors distributed across all bits of key data (as opposed to the most- or least significant bits only), [3] proposed a method different from previously published work. Given public modulus N and $p'$ and $q'$ recovered from memory, the original $p$ and $q$ can be reconstructed iteratively from the least significant bits. Under conditions of perfect asymmetric decay with probability $\delta$, the $i_{th}$ bits of p and q are uniquely determined by the $i_{th}$ bit of N and the attacker's guesses for the i-1 least significant bits of p and q (except when $p'_i$ and $q'_i$ are both in ground state) yielding a branching process of degree $\frac{(3+\delta)^2}{8}$. Under conditions of imperfect asymmetric decay, the estimated probabilities can be used to weight the branches at each given bit guess. For 2048-bit keys, the algorithm achieved median reconstruction times of 4.5 seconds where $\delta = 4\%$ and 2.5 minutes where $\delta = 6\%$, while for 1024-bit keys it achieved a median reconstruction time of 1 minute where $\delta = 10\%$. For higher error rates, it is proposed to reconstruct only the first $\frac{n}{4}$ bits using this technique and use the techniques described in [34,35,36,37] to efficiently reconstruct the full key.

Since the publication of the paper by Halderman et al. [3] several papers (the full contents of which are outside the scope of this paper) seeking to tackle the problem of RSA key reconstruction under conditions of partial key exposure with a focus on CBAs have been published [39,40,41,42,43,44], mostly relying on binary-tree-search and pruning-based techniques. However not all of the proposed methods are equally well-applicable to CBA scenarios as Halderman et al. [3] observe that in a realistic CBA scenario the asymmetric decay is likely to be imperfect. Kunihiro et al. [42] describe two effects introducing noise in to-be-reconstructed key material: *erasures*

(bits equal to the decay ground state that can as such not be known to be in tact or decayed, denoted as δ) and *errors* (bits flipping in the opposite direction, denoted as ε). Most methods for RSA key reconstruction, however, consider only one of two cases (eg. only erasures or only errors) and are incapable of dealing with keys (called *noisy secret keys* by Kunihiro et al.) suffering from both. Paterson et al. [44] add to this the observation that methods relying on the assumption that a given fraction (consisting of a mixture of 0s and 1s) of the RSA private key is known [34,35,36,37,39] with certainty, are hindered by the fact that, in a CBA scenario (even under unrealistic conditions of perfect asymmetric decay), only either all 0s or all 1s are known with certainty. In addition, methods relying on the assumption of (near) symmetric decay (ie. $\delta_0 = \delta_1$) [40,41] overlook the fact that in a CBA scenario $\delta_1$ will virtually always be relatively small while $\delta_0$ will virtually always be several times larger than $\delta_1$, as such rendering methods built upon symmetric decay assumptions unreliable in real-world CBA scenarios. To our knowledge, only the works [42,43,44] are capable of properly and efficiently dealing with realistic CBA scenarios.

Kunihiro et al. [42] present a polynomial time algorithm, by combining and optimizing the methods of Heninger et al. [39] and Henecka et al. [40], for noisy secret key reconstruction using noise model $0 \leq \varepsilon < \frac{1}{2}$, $0 \leq \delta < 1$ and $0 \leq \varepsilon + \delta < 1$. The proposed algorithm can handle special cases (ie. the perfect asymmetric decay and symmetric decay assumed by [39] and [40] respectively) achieving the upper bounds of its composite algorithms in these cases and outperforming them in general. The method can reconstruct 1024-bit AES keys in polynomial time provided that $1 - \delta - 2\varepsilon \geq \sqrt{\frac{2(1-\delta)ln(2)}{m}}$ (where m is the number of secret parameters (eg. *p, q, d,* etc.) involved), with success probability over 91% and running time below 2.24 seconds for $\delta_0$ up to 0.6 and $\delta_1$ up to 0.01.

Paterson et al. [44] recast the previous work on RSA key reconstruction under conditions of partial key exposure as a coding theoretical problem, designing a method capable of dealing with a realistic (ie. imperfect asymmetric decay) CBA scenario. The proposed algorithm is a modification of the one proposed by [39] but uses the Maximum Likelihood (ML) [45] statistic instead of the Hamming metric for candidate selection. At each stage the algorithm selects *L* of the candidates (generated using Hensel lifting [46] on the maintained list) with the highest ML value to be passed as the list to the next stage (with the initial list being constructed from the recovered decayed key material), testing the final list of candidates with encryption and decryption to obtain a single private key candidate. In addition to the proposed algorithm, Paterson et al. also propose a method to derive upper bounds (in terms of $\delta_0$) on the performance of algorithms used for noisy RSA key reconstruction. The proposed method matches or outperforms those of [39,40] (in the applicable cases) and is capable of achieving non-zero success probabilities for $\delta_0$ as high as 0.61 (the theoretical upper bound upon which is 0.658) given $\delta_1 = 0.001$. Success probabilities of at least 66% are achieved for $\delta_0$ up to 0.5, with running times always below 42.732 seconds. In addition, when private keys of the forms (*p, q, d*) and (*p, q*) are concerned (under identical $\delta_1$), the method achieves non-zero success probabilities for $\delta_0$ up to 0.43 and 0.26 respectively (with theoretical upper bounds 0.479 and 0.298 respectively). For these cases success probabilities of at least 55% and 68% are achieved for $\delta_0$ up to 0.35 and 0.15 with running times always below 108.523 and 9.492 seconds, respectively.

In contrast with search tree- [39] or lattice-based [35] approaches, Patsakis [43] proposes the use of SAT solvers for the purpose of RSA key reconstruction. The problem is cast as a Boolean satisfiability problem linking the bits of the private to the public RSA key, filling in the bits known through partial key exposure. While performing not as efficiently as the aforementioned approaches, this method is capable of reconstructing 1024-bit RSA keys within 40 seconds with only 38% known bits of *p*, *q* and *d*. It is further noted that implementation improvements as well as further research in the usage of SAT solvers for RSA key reconstruction would probably yield even better results.

In addition to the above approaches showing the feasibility of RSA key reconstruction in a realistic CBA scenario, it is noted that the approaches of [42] and [44] can be adapted to first reconstruct either the RSA key's Least- or Most Significant Bytes before passing on the result to methods optimized for dealing with reconstructing keys given such partial exposure (but which are generally incapable of dealing with realistic CBA scenarios) such as [34,35,36,37,39].

### 2.4 Key identification

In order to obtain (and reconstruct) encryption keys from dumped memory images, an attacker has to be able to locate them first. The problem of identifying cryptographic key information from acquired memory images has featured prominently in the field of digital forensics for quite some time, as outlined by [47,48], and has seen many different approaches over the years:

1. *Brute-force*: A straightforward, if naïve, approach involves testing every possible byte sequence as a potential key in a linear fashion [49] or using the memory image as a dictionary [50]. Despite being easily generalizable such approaches only work properly when cryptographic keys are stored directly and contiguously in memory. In addition they quickly become infeasible for large memory images and require an accurate, error-free memory image.

2. *Mathematical properties*: Techniques seeking to identify cryptographic keys on the basis of their mathematical properties (such as the relation between known public keys and the searched-for private key in the case of RSA or their high entropy) have been proposed by Shamir et al. [51]. Such methods have the benefit of not requiring the in-memory presence of any special format or key schedule but are susceptible to a fairly high number of false posi-tives [3], especially in the context of decayed memory images as well as being generally inapplicable to symmetric keys [50].

3. *Identifying encoding formats*: Such approaches, as proposed by [3,52,53], avoid searching for the key material directly and instead seek to identify them by meta-data related to their encoding formats, such as finding RSA keys by looking for ASN.1 prefixes [54] of the DER-encoded [55] private keys [60]. The methods of [52,53], however, have been noted to be quite prone to false positives [3] and generally inapplicable to symmetric keys [50] and as such [3] proposes to either search for known fields of the ASN.1 object (such as the public modulus) or look for the DER encoded RSA version number (instead of RSA Object Identifiers) and the subsequent DER encoding of the next field a few bytes further.

4. *Identifying data structures*: These techniques, such as those proposed by Pettersson [10] and Walters et al. [56], seek to identify keys by their container data structures. The method in [10] involves testing memory locations for their plausibility of holding key data based on the likely values for surrounding variables. It is noted [3], however, that this method requires an attacker to manually derive search heuristics for each encryption application as well as not being very robust in the face of memory errors.

5. *Identifying key schedules*: [3] proposed a fully automated technique for locating encryption keys in memory images regardless of the presence of errors. The proposed technique identifies the key schedule instead of the master key itself and searches for the schedule by looking for memory blocks which (almost) satisfy its combinatorial properties, allowing an attacker to recover partial key schedules (such as those overwritten during memory reallocation) as well as avoiding having to reverse-engineer the encryption software in order to identify the particular data structures used. [59] significantly improves upon the work of [3] by implementing the proposed AES key identification method as an FPGA, outperforming a software implementation from 10 up to 75 times when tested against real memory contents and from 25 up to 205 times when tested against synthetic random data, depending on the chosen error thresholds (since the approach of [3] discards candidates exceeding a given threshold) and the number of parallel FPGA kernels.

6. *Combined approach*: [48] proposed an approach combining several of the aforementioned methods, implemented in the proof-of-concept tool Interrogate [57] capable of searching for RSA, AES (128-, 192- or 256-bit), Serpent [61] and Twofish [62] (256-bit) keys. An additional contribution of [48] is proposing a technique to reconstruct process virtual address space from the acquired memory image using the so-called Page Directory Base (PDB) [58], which significantly improves search time (in the face of keys distributed non-contagiously throughout memory due to paging).

## 3 Practical applicability

In this section, we will present a brief overview of the results of various research efforts into the practical applicability of CBAs against a variety of systems and cryptographic solutions. While some works [48] have addressed several different cryptographic software classes (such as Full Disk Encryption and Virtual Disk Encryption), for brevity's sake we will only summarize the results of experiments against systems offering FDE (and optionally VDE) functionality.

### 3.1 PCs

The primary target of most CBAs will be various types of PCs, especially portable ones such as laptops or netbooks. As noted by [48], an attacker performing a CBA can be faced with a machine in a large number of different system states (which are reduced to 8 'archetypical' states in the paper). It was found that [48], generally speaking, all tested FDE systems were fully vulnerable (ie. recovery with 100% success rates) in the live, screensaver and logged out states, less vulnerable (with 29% success rate) in reboot state and not vulnerable in boot state. They also found VDE key management routines regarding hibernation-based dismounting to be inadequate, allowing for recovery rates of up to 44% from the hibernation file. Furthermore, it is noted that successful identification and reconstruction of cryptographic key material is dependent upon the system state at attack time with CBAs against live, screensaver and logged out system states being highly successful and turned off devices yielding far worse results. The authors thus conclude that target system state at the moment of acquisition and successful identification of cryptographic systems and hardware in use by the target are vital to CBA success. Provided that these conditions are met, the chances of a successful CBA attack are considered sufficiently high by the authors to consider it a valuable and realistically applicable digital forensic tool.

The experiments conducted by [64] against a variety of system configurations (with memory sizes ranging from 256MB to 24GB) running an instance of TrueCrypt configured for VDE purposes yielded some contradictory results with [3]. It was noted that not all memory modules tested exhibit signs of data remnance and that those which do sometimes wouldn't when used in certain computer systems, possibly owing to the role motherboard residual capacitance plays in the maintenance of electrical charges of the memory module cells. Since there is no available exhaustive listing detailing data remnance properties for various memory modules and their combination with various motherboards, this underlines the importance of target system knowledge by an attacker pointed out by [48]. As such [64] recomms CBAs only as a last resort digital forensic tool, favoring more forensically sound and reliable techniques (such as DMA attacks [65]) instead.

Gruhn et al. [66] verified the claims made by [3] against a variety of memory modules. The conducted tests confirmed the practicality of CBAs against DDR1 and DDR2 memory modules under a variety of setups even though deviations with re-gards to temperatures and error rates were found. In addition, the correlation between memory module temperatures and data remnance was confirmed empirically as well as confirming that a CBA involving transferring (significantly cooled) memory modules between target and attacker systems (see section 2.2) was feasible in practice with an error rate low enough (between 1 and 5%) to allow for efficient key reconstruction and the additional benefit (for an attacker) of bypassing all software-based countermeasures. A new result, however, was the finding that modern DDR3 memory modules (not covered in [3]) do not exhibit data remnance under cold boot conditions (even though a warm boot attack scenario, which could be prevented with a BIOS boot lock, did prove successful), regardless of temperature, possibly owing to either the combination of lower voltages, higher integration density and hence lower RAM cell charges or specifics related to DDR3 memory controllers. If the later proves to be the case, the authors argue, it is possible that an attacker machine with modified DDR3 controllers could allow for CBAs against such memory modules.

Thus we can conclude that, while advice regarding their forensic reliability varies, CBAs are a practically feasible class of attacks (for a wide range of memory modules and system setups) with a high success rate (given adequate pre-existing attacker knowledge of the target system) and as such pose both a privacy and security risk and an opportunity for practitioners of digital forensics.

### 3.2   Smartphones

With the proliferation of smartphones and the increasingly diffused nature of sensitive data storage [67], smartphones present an interesting target to all kinds of attackers (criminal and forensic investigator alike) seeking to tap into the wealth of information held by these devices. Müller et al. [68] documented the feasibility of CBAs against ARM-based portable devices such as Android smartphones. With the release of Google's Android 4.0 operating system in 2011, smartphone users were able to use FDE to protect the contents of the data in their user partitions. Android 4.0's FDE solution (which is disabled by default) is based on dm-crypt [69] and encrypts user partitions mounted at /data (instead of the whole disk) using AES in dm-crypt's *aes-cbc-essiv:sha256* mode using 128-bit keys. Data is encrypted using a primary '*Data Encryption Key*' (DEK) which is subsequently encrypted using a secondary '*Key Encryption Key*' (KEK) derived from a user-defined PIN or password in order to avoid necessary re-encryption of all data in case the latter are changed. The user-defined PIN or password in question is the one used for so-called screenlock [70] functionality and as such Android's encryption functionality is not available when screen locking mechanisms such as patternlocks (which can be easily bypassed with so-called smudge attacks [71]) or facial recognition (which can be easily bypass by holding up a picture of the owner) are used. PINs are comprised of 4 to 16 numeric characters while passwords are comprised of 4 to 16 alphanumeric characters with at least one letter, with 4-digit PINs being the most widely used screenlock protection in use (even as 57% of all smartphones aren't protected by any security mechanism at all).

Müller et al. [68] presented a tool called FROST (developed on the basis of a CBA against Galaxy Nexus smartphones specifically, but generalizable to other models) to perform CBAs against Android smartphones with encrypted user partitions. FROST follows the general CBA procedure outlined in section 2, with the difference lying mainly in the acquisition of the RAM memory dump. Since smartphones cannot be booted from USB or other external media, as is the case for PCs, the existence of the *recovery* partition (comparable with a rescue system for regular PCs as it allows basic operations without fully booting into Android) is abused to flash the prepared FROST image onto. It is observed that there are generally two scenarios that an attacker can encounter:

1. *Unlocked bootloader*: On devices with an unlocked bootloader, FROST is capable of cryptographic key recovery from RAM or PIN-bruteforcing and immediate subsequent on-phone user partition decryption using the obtained key. FROST offers two key search modes: *quick* and *full*, with the former being optimized for Galaxy Nexus devices and only scanning for AES keys within a certain address range (doing so within seconds, though with a lower success rate on other devices) and the latter searching the entire address space using a sliding window mechanism looking at every physical memory page twice (in order to account for AES keys possibly being split over multiple memory pages), doing so in between roughly 3 and 8 minutes. It is further noted that 4-digit PINs can be cracked within at most ~3 minutes and that extension to up to 7-digit PINs would be feasible (taking about 25 hours).

2. *Locked bootloader*: Unfortunately for an attacker, bootloader unlocking generally (with a few models as exceptions) wipes userdata and cache partitions which renders cryptographic key recovery pointless (as the encrypted data is gone) though still possible. However, it is noted that there is a relatively large amount of Galaxy Nexus devices (and other smartphones as well) with unlocked bootloaders. In addition, it is still possible for an attacker to recover a wealth of sensitive information from the memory of smartphones with locked bootloaders using a CBA such as personal photos, visited websites, e-mail and instant messaging logs, contact lists, calendar entries and WLAN credentials.

Before performing a CBA against the target smartphone, an attacker has to ensure the device has enough power because after power loss the only option left is PIN-bruteforcing. After ensuring adequate power, the device has to be cooled down (putting the phone in a water-condensation resistant bag for a period of 60 minutes in a -15°C freezer was found to be sufficient) to increase the attack time window (since an error rate of 50% was already noted at temperatures of only 25°C, while only 5% was observed at temperatures of 10°C). Assuming the bootloader is unlocked, the CBA itself is then performed by removing and quickly re-inserting the phone battery while holding the power button before removal, so as to ensure quick booting upon battery re-insertion. Now that the risk of memory decay is averted, so-called *Fastboot* mode is entered, the FROST image is flashed onto the phone from a connected PC and the phone is booted into the recovery mode which now contains the FROST image.

## 4   Counter-measures

Below we will discuss a variety of counter-measures, both software- and hardware-based, that have been proposed to mitigate memory-acquisition attacks in general and CBAs in particular with various degrees of effectiveness.

### 4.1   Software-based

1. *Scrubbing memory* [3]: A primary countermeasure would be avoiding the storage of cryptographic key material in memory whenever possible and erase it as soon as it is no longer being used. Software ought to (securely) overwrite key material when it is no longer needed and prevent its paging to the disk and boot procedures ought to perform wiping of any remnant data as well. However, this does not apply to keys that need to be in use such as those used by mounted encrypted disks and an attacker could still perform a memory module transference-based CBA [66]. An example of the use of memory scrubbing mitigation is DEADBOLT [72], which seeks to achieve protection against CBAs for Android smartphones by (upon screenlock) un-mounting the encrypted userdata partition and securely overwriting the key stored in RAM as well as possible sensitive plaintext information.

2. *Limiting boot options* [3]: Many CBA scenarios involve booting a system from removable media or other non-standard booting options. Limiting the allowed boot options (eg. to disk-only) and setting an administrative password offers some protection, though an attacker might still be able to swap out the system's drive for a custom one, reset flash memory to re-enable non-standard booting or simply transfer memory modules [66].

3. *Safe system suspension* [3]: System suspension (eg. entering '*sleep*' mode) has been shown to be vulnerable to CBAs, even if the system enters screen-lock mode upon wake (since the attacker could simply cold boot the machine after waking). Similarly, system suspension to the disk (eg. entering '*hibernation*' mode) is also vulnerable unless an externally-contained secret is required for system resumption. An effective countermeasure would be simply powering off the machine and guarding it for a couple of minutes until memory contents are properly decayed before leaving. However, this is quite inconvenient for end-users (who are hence less likely to adopt it) and as such safe system suspension might be better achieved by requiring an external se-cret (such as a password) for wakeup and memory encryption (see section 4.2) using a key derived from this secret.

4. *Avoiding pre-computation* [3]: As discussed in sections 2.3 and 2.4, precomputation makes key identification and reconstruction far easier. So while pre-computation might increase performance, avoiding it altogether or making compromises in the form of time-limited caching of precomputed values reduces vulnerability to successful CBAs.

5. *Temperature detection* [66]: The usage of temperature sensors [73] has been proposed as a way to complicate, but not fully prevent, CBAs. As many modern motherboards have temperature sensors built-in for heat control (including for the RAM zone), software could erase memory contents upon detecting a sudden temperature drop within the RAM zone. Since sensors are infused into the motherboard, the risk of tampering is mitigated. However, an attacker is still able to power up an attacker-controlled system B without RAM (which, though left in a failure state, keeps refreshing any inserted memory module) and rapidly remove the desired memory module from target system A (without using cooling) and inserting it in B, leaving the attacker with a regular CBA scenario on system B. Since no cooling is used, the CBA on B is performed under conditions of more serious decay though test results have shown that 90 to 99% of memory contents could be recovered correctly (though its proneness to mishaps means that scenarios with recovery of only 73% of memory contents can occur as well).

6. *Boot block defense* [66]: Another proposed countermeasure aimed at x86 architecture machines is the so-called 0x7c00 (or *boot block*) method. On x86 machines 0x7c00 is the memory address to which the BIOS loads the Master Boot Record (MBR). Cryptographic key material could then be placed in this 512-byte sized memory region such that any reboot will automatically overwrite it. However, an attacker could still transplant the target memory module to a system with two memory slots where slot covering the 0x7c00 memory region is filled with a dummy memory module while the slot covering the upper address space holds the target memory module.

7. *CPU-bound cryptography*: Given that RAM is the acquisition target of CBAs and other memory-based side-channel attacks (such as DMA attacks), one proposed category of solutions consists of kernel modifications which store sensitive cryptographic key material securely outside RAM in such a way that they aren't easily accessed by applications running with regular privileges and are lost as soon as the computer restarts. There exist various proposals for such an approach:

    i. *Symmetric*: TRESOR [74] is a Linux kernel patch implementing AES and the associated key management purely within the CPU, avoiding

storing sensitive information in RAM. To this end, it uses the CPU debug registers [75] to store the keys, as the contents of these registers are irrecoverably lost upon any power-off. In addition, TRESOR uses Intel's AES-NI [76] instruction extensions (implementing most of AES in hardware) in order to optimize code length and performance so that it matches (and in some cases even outperforms) the performance of AES implementations using RAM. AES-NI instruction operands are always SSE [77] registers and as such cryptographic state and key information passed to these instructions is fully CPU-bound during execution as well as during passive storage. Since CPU registers are often swapped to RAM (due to user space scheduling), a naïve CPU-bound AES implementation might still have sensitive information end up in memory. In order to mitigate this, TRESOR disables interrupts (eg. it is run atomically) while cryptographic operations are being performed as well as blocking access to the debug registers by non-ring 0 processes. The only time sensitive material enters RAM is for a brief period when a system users enters their password during startup right before it is transferred to the debug registers, after which the associated memory is securely wiped. Since key schedules cannot be kept in-memory (due to security reasons) nor in CPU registers (due to lack of space), TRESOR recomputes round keys on-the-fly for each input block (which is processed within an atomic section where round keys can be safely stored in the SSE registers). Luckily, the performance drawbacks of round key recomputation are mitigated by the usage of specialized AES-NI instructions. As a secure side-effect of its implementation, TRESOR complicates (if not mitigates) AES key reconstruction due to its elimination of persistent key schedule storage and complicates DMA attacks by eliminating sensitive material from memory. It is noted, however, that it is theoretically possible to read out CPU registers on a running machine using an oscilloscope by measurement of the electromagnetic field around the CPU, even though there seem to be no such successfully documented attacks. With TreVisor [78], TRESOR's encryption capabilities were added to BitVisor [79], a light-weight hypervisor implementing a variety of security features (such as FDE and protection against certain DMA attacks). This way TRESOR is made OS-independent by moving it from the operating system to an underlying hypervisor, thus offering protection against CBAs regardless of operating system. Finally, TRESOR can be considered the conceptual father to ARMORED [80], which is a dm-crypt module that offers CPU-bound encryption for ARM-based Android smartphone devices in order to mitigate (among other things) CBAs against them.

Loop-Amnesia [81] is a Linux kernel patch for CPU-bound AES encryption (based on Loop-AES [82]) conceptually similar to TRESOR with the primary differences being Loop-Amnesia's usage of Machine Spe-cific Registers (MSRs) [83] for key storage, its (as of yet) lack for AES-NI support, limitation to 128-bit keys only and support for multiple disk encryption keys (as opposed to TRESOR's limitation to a single key and hence a single encrypted disk) by storing them securely in RAM after encrypting them with the primary, CPU-bound master key, thus offering multiple encrypted disks (allowing mitigations against watermarking attacks as offered by Loop-AES) at the cost of worse performance and lower overall key length.

However, as [84] has shown, an attacker capable of performing a DMA attack is able to bypass CPU-bound encryption solutions. The OS-independent attack is based on the observation that solutions like TRESOR make the assumption that while ring 3 processes are to be shut off from access to cryptographic keys, ring 0 processes are not and that the latter will not leak key material to RAM as well as the fact that DMA attacks can read as well as write to memory. The attack itself has an attacker gain physical access to the target system, attach a device to a DMA-bus (eg. FireWire) and recover physical memory contents for subsequent analysis. The attack devices injects a payload into target memory at the adequate locations such that it executes within ring 0. The payload (now running in the kernel) then proceeds to copy the CPU-bound encryption key from the registers it is kept in to a given location in RAM for subsequent DMA-based extraction by the device. The authors note that the attack, which was confirmed to be effective against TRESOR and suspected to be effective against Loop-Amnesia as well, could be mitigated through a variety of measures such as disabling DMA, DMA device whitelisting, Hardware-based FDE (see section 4.2) or the usage of an IOMMU [86].

ii. *Asymmetric*: PRIME [87] is a Linux kernel patch that extends the notion of CPU-bound encryption to support RSA and hence closes the gap for asymmetric CPU-bound encryption support. It does this by symmetrically encrypting all private RSA parameters for the supported 2048-bit keys with AES (using the AES-NI instructions) before storing them in RAM and decrypting them only within CPU registers as well as implementing RSA's modular exponentiation fully on the CPU (using an optimized version of Montgomery reduction [88]) such that no state information is leaked to RAM. PRIME only supports private RSA operations (as public RSA operations are non-critical with regards to the posed problem) and has a performance drawback of being roughly 10 times slower than regular RSA implementations. As a secure side-effect of its implementation PRIME complicates potential RSA key reconstruction by not encoding its private keys in the fashion described in section 2.4. COPKER [89] is a Linux kernel patch for CPU-bound private RSA operations similar to PRIME with the difference of having a larger cache-size and thus support for longer private keys and more efficient algorithms (eg. RSA with Chinese Remainder Theorem [90]).

8. *Leakage-resilient cryptography* [3]: The problem of designing cryptosystems resilient to partial key exposure (which is effectively the basis of key reconstruction efforts from memory images acquired through CBA) extends beyond CBAs [91,92,93] and the exact specifics fall outside the scope of this paper. Certain proposals [91], however, have been proposed in a CBA context [3]. The approach in [91] proposes to render a cryptographic key K, not currently in use but necessary at a later time, resilient to partial key exposure by storing it as K $\oplus$ H(R), where R is a *b*-bit sized buffer of random data and H a hash function (eg. SHA-256). Given a CBA scenario where *d* bits in the buffer are flipped (and assuming the use of a strong hash function) the attacker has a search space of $\binom{\frac{b}{2}+d}{d}$ to determine which were flipped given that $\frac{b}{2}$ bits could have decayed. If *b* is sufficiently large this will be infeasible even for relatively small *d*. For normal operations, when the key is required again recomputation given R is trivial and fast. Though interesting developments have been made in this field, it is noted [81] that the fact that CBA memory acquisition can yield images with error rates under 1% means that partial key exposure resilient cryptosystems by themselves aren't sufficient as a protection measure as any definitive countermeasure has to be able to withstand the possibility of full key exposure to RAM.

### 4.2 Hardware-based

1. *Memory encryption*: The successful recovery of sensitive information from a memory dump acquired by an attacker could be mitigated by the usage of memory encryption (varying from One-Time-Pad schemes [94] to AES), where memory contents are encrypted upon writing to RAM and decrypted upon reading. Initially designed around questions regarding intellectual property theft and assurance of memory integrity, (full or partial) memory encryption can be roughly divided into three areas: hardware-based modifications, software-based modifications or specialized industrial applications. Of these, only the former two are applicable to CBA mitigations for general purpose computing devices. There are a variety of memory encryption approaches [95] and though there exist both proposed software- and hardware-based memory encryption solutions, it has been noted that the most mature, efficient and secure proposals are hardware-based in nature. There are several reasons for this: First of all, the increased overhead associated with software-based memory encryption solutions (largely resulting from them being implemented on top of the existing operating system) results in its lowered likelihood of widespread adoption. Additionally, in order to compensate for this overhead, many software-based memory encryption approaches employ a partial encryption model where a set of 'secure components' (such as cryp-tographic application processes) is identified and only their associated memory regions are encrypted. Apart from the non-trivial nature of identifying 'secure components', there is always the possibility of memory contents and sensitive information being exchanged with other, non-secure, components which use unencrypted memory regions. Finally, software-based solutions make a series of assumptions (eg. that the kernel is secure to begin with) that hardware-based solutions do not make. According to the authors of [95] the most secure memory encryption approach is Aegis [96] which includes both hardware and software components, integrity mechanisms, full memory encryption using AES and an FPGA-implemented prototype. In practice, however, there seems to be only one commercial product (PrivateCore's vCage [97]) offering memory-encryption functionality with other proposals being academic works and their associated proof-of-concepts. To our knowledge, there exists no literature evaluating the practical effectiveness of memory encryption technology against CBAs.

2. *Hardware-based FDE* [98]*:* Hardware-based FDE (sometimes called *self-encrypting drives (SED)*) performs cryptographic operations using specialized chips in the hard disk controller instead of using system CPU and RAM, thus eliminating memory as an attack vector (and hence mitigates CBAs). Generally speaking, this approach involves the usage of a write-only key register internal to the disk controller to which software can write (but not read from) an encryption key, which is then used to encrypt data before writing it to the disk and decrypt data before reading it from the disk. As such, cryptographic functionality is fully contained within the disk controller and the associated key isn't stored in either RAM or the CPU, with the additional benefit of having no performance degradation. A drawback, compared to software-based FDE, is that the latter is usually cheaper, disk-vendor independent, more configurable and quickly and easily employable on existing system setups. However, it was found [99] that theoretically secure SEDs are often vulnerable to a variety of (new and adapted) attacks in practice. For example, SEDs do not detect whether their data connection is cut as long as power remains on and hence only get locked when power is cut. Given a running system with an unlocked SED, an attacker could thus perform a so-called *hot-plug attack* by unplugging a target SED's data connection (while ensuring power remains connected) and plug it back into its own machine and proceed to read out the still unlocked data. In addition it was found possible to reboot running systems with unlocked SEDs (since the SED doesn't realize it is getting rebooted as power to it remains on) into a different (live) operating system which could then proceed to simply mount the unlocked disk. Both attacks are trivial and far more reliable than regular CBAs, making most SEDs less secure than software-based FDE solutions. Security of this approach thus depends on at least ensuring that the key register is erased and the drive locked upon power loss or when it is attached to another system (even if no power loss occurs during the move).

3. *Architectural changes* [3]: Designing RAM which exhibits faster decay might be a solution to the CBA problem, though obviously there is the tension between faster decay and ensuring decay probability between refresh intervals remains low. As discussed in section 3.1, DDR3 might be a candidate for such a solution though as of yet it remains to be determined whether the low remnance exhibited is intrinsic to DDR3's design or related to its controllers.

4. *Physical mitigations* [3]: Physical protection of RAM modules (through locking them in place with soldering, encasing them in epoxy, etc.) might be a simple way to mitigate or at least complicate CBAs, at the cost of reducing system modification flexibility.

# 5 Conclusion

In conclusion, the data remnance effects exhibited by many RAM memory modules are a viable and realistic attack vector for an attacker seeking to extract sensitive information contained in-memory (such as cryptographic keys to circumvent FDE systems) through CBA. The decay to which the acquired information is subject can be sufficiently minimized through physical cooling of the memory modules, so that a variety of approaches for the identification and reconstruction of cryptographic key material under conditions of decay can be feasibly executed within a realistic timeframe. CBAs have been shown to be practically applicable to both PCs and smartphones given that the attacker encounters these devices in a particular system state. However, the fact that certain memory modules (such as DDR3 RAM) exhibit no signs of data remnance (though it still has to be determined whether this is intrinsic to their design) and the lack of an exhaustive study of which memory modules in which system setup are susceptible and which aren't, combined with the necessity for an attacker to determine the target system's memory module type on forehand puts limitations on the applicability of CBAs to the field of digital forensics, as the possibility of data-loss or corruption due to a botched CBA presents technical as well as judicial difficulties. In such cases, it is generally recommended other, less delicate, memory acquisition side-channel attacks (such as FireWire DMA attacks) are used instead when possible. From the variety of proposed counter-measures to CBAs, CPU-bound cryptography seems to be the most promising software-based solution, though hardware-based solutions seem more resilient in general (memory encryption in particular) even though very little robust solutions seem to be openly available and a solution presenting combined counter-measures protecting against the entirety of the memory-acquisition attacks spectrum (such as CBAs and DMAs) remains an open problem.

# References

1. http://en.wikipedia.org/wiki/Side_channel_attack. Last visited 24-06-2014
2. http://en.wikipedia.org/wiki/Cold_boot_attack. Last visited 24-06-2014
3. J. Halderman et al. *Lest We Remember: Cold Boot Attacks on Encryption Keys*. In Proc. 17th USENIX Security Symposium, 2008
4. http://en.wikipedia.org/wiki/Data_remanence. Last visited 24-06-2014
5. http://en.wikipedia.org/wiki/Disk_encryption#Full_disk_encryption. Last visited 24-06-2014
6. E. Casey et al. *The Impact of Full Disk Encryption on Digital Forensics*. In ACM SIGOPS Operating Systems Review, 2008
7. http://en.wikipedia.org/wiki/TrueCrypt. Last visited 24-06-2014
8. http://en.wikipedia.org/wiki/BitLocker. Last visited 24-06-2014
9. http://en.wikipedia.org/wiki/Dynamic_random-access_memory. Last visited 24-06-2014
10. T. Pettersson. *Cryptographic key recovery from Linux memory dumps*. In Chaos Communication Camp 2007
11. P. Wyns et al. *Low-temperature operation of silicon dynamic random-access memories*. In IEEE Transactions on Electron Devices Vol 36, Issue 8, 1989
12. P. Gutmann. *Data remanence in semiconductor devices*. In Proc. 10th USENIX Security Symposium, 2001
13. http://en.wikipedia.org/wiki/Power-on_self-test. Last visited 24-06-2014
14. http://en.wikipedia.org/wiki/Key_schedule. Last visited 24-06-2014
15. http://en.wikipedia.org/wiki/Error-correcting_code. Last visited 24-06-2014
16. http://en.wikipedia.org/wiki/Data_Encryption_Standard. Last visited 24-06-2014
17. http://en.wikipedia.org/wiki/Repetition_code. Last visited 24-06-2014
18. http://en.wikipedia.org/wiki/Advanced_Encryption_Standard. Last visited 24-06-2014
19. A. Tsow. *An Improved Recovery Algorithm for Decayed AES Key Schedule Images*. In Lecture Notes in Computer Science Volume 5867, 2009
20. http://en.wikipedia.org/wiki/Branch_and_bound. Last visited 24-06-2014
21. H. Riebler et al. *Reconstructing AES Key Schedules from Decayed Memory with FPGAs*. In Proc. Int. Symp. Field-Programmable Custom Computing Machines, 2014
22. http://en.wikipedia.org/wiki/Field-programmable_gate_array. Last visited 24-06-2014
23. http://en.wikipedia.org/wiki/Integer_programming. Last visited 24-06-2014
24. M. Albrecht et al. *Cold Boot Key Recovery by Solving Polynomial Systems with Noise*. In Lecture Notes in Computer Science Volume 6715, 2011
25. http://en.wikipedia.org/wiki/SCIP_(optimization_software) . Last visited 24-06-2014

26. Z. Huang et al. *A New Method for Solving Polynomial Systems with Noise over F2 and Its Applications in Cold Boot Key Recovery*. In Lecture Notes in Computer Science Volume 7707, 2013
27. F. Chai et al. *A Characteristic Set Method for Solving Boolean Equations and Applications in Cryptanalysis of Stream Ciphers*. In Journal of Systems Science and Complexity, Volume 21, Issue 2, 2008
28. A. Kamal et al. *Applications of SAT Solvers to AES key Recovery from Decayed Key Schedule Images*. In 4th Int. Conf. on Emerging Security Information Systems and Technologies (SECURWARE), 2010
29. http://en.wikipedia.org/wiki/Boolean_satisfiability_problem. Last visited 24-06-2014
30. M. Soos et al. *Extending SAT Solvers to Cryptographic Problems*. In Lecture Notes in Computer Science Volume 5584, 2009
31. https://github.com/msoos/cryptominisat. Last visited 24-06-2014
32. X. Liao et al. *Using MaxSAT to Correct Errors in AES Key Schedule Images*. In IEEE 25$^{th}$ Int. Conf. on Tools with Artificial Intelligence (ICTAI), 2013
33. http://en.wikipedia.org/wiki/RSA_(cryptosystem) . Last visited 24-06-2014
34. R. Rivest et al. *Efficient factoring based on partial information*. In Proc. of a workshop on the theory and application of cryptographic techniques on Advances in cryptology – EUROCRYPT, 1985
35. D. Coppersmith. *Small Solutions to Polynomial Equations, and Low Exponent RSA Vulnerabilities*. In Journal of Cryptology, Volume 10, Issue 4, 1997
36. D. Boneh et al. *Exposing an RSA Private Key Given a Small Fraction of Its Bits*. 1998
37. J. Blömer et al. *New Partial Key Exposure Attacks on RSA*. In Lecture Notes in Computer Science Volume 2729, 2003
38. http://en.wikipedia.org/wiki/Lattice_reduction. Last visited 24-06-2014
39. N. Heninger et al. *Reconstructing RSA Private Keys from Random Key Bits*. In Proceedings of the 29th Annual International Cryptology Conference on Advances in Cryptology, 2009
40. W. Henecka et al. *Correcting Errors in RSA Private Keys*. In Lecture Notes in Computer Science Volume 6223, 2010
41. S. Sarkar et al. *Error Correction of Partially Exposed RSA Private Keys from MSB Side*. In Lecture Notes in Computer Science Volume 8303, 2013
42. N. Kunihiro. *Recovering RSA Secret Keys from Noisy Key Bits with Erasures and Errors*. In Lecture Notes in Computer Science Volume 7778, 2013
43. C. Patsakis. *RSA private key reconstruction from random bits using SAT solvers*. 2013
44. K. Paterson et al. *A Coding-Theoretic Approach to Recovering Noisy RSA Keys*. In Lecture Notes in Computer Science Volume 7658, 2012
45. http://en.wikipedia.org/wiki/Maximum_likelihood. Last visited 24-06-2014
46. http://en.wikipedia.org/wiki/Hensel's_lemma#Hensel_Lifting. Last visited 24-06-2014

47. S. Vömel et al. *A survey of main memory acquisition and analysis techniques for the windows operating system*. In Digital Investigation: The International Journal of Digital Forensics & Incident Response, Volume 8, Issue 1, 2011
48. C. Maartmann-Moe et al. *The persistence of memory: Forensic identification and extraction of cryptographic keys*. In Digital Investigation: The International Journal of Digital Forensics & Incident Response, Volume 6, 2009
49. C. Hargreaves et al. *Recovery of Encryption Keys from Memory Using a Linear Scan*, Proceedings of the Third International Conference on Availability, Reliability and Security, 2008
50. B. Kaplan. *RAM is Key - Extracting Disk Encryption Keys From Volatile Memory*. 2007
51. A. Shamir et al. *Playing 'Hide and Seek' with Stored Keys*. In Lecture Notes in Computer Science Volume 1648, 1999
52. T. Klein. *All your private keys are belong to us*. 2006
53. http://mtso.squarespace.com/chargen/2006/1/25/recover-a-private-key-from-process-memory.html. Last visited 24-06-2014
54. http://en.wikipedia.org/wiki/Abstract_Syntax_Notation_One. Last visited 24-06-2014
55. http://en.wikipedia.org/wiki/X.690#DER_encoding. Last visited 24-06-2014
56. A. Walters et al. *Volatools: integrating volatile memory forensics into the digital investigation process*. Black Hat DC, 2007
57. http://sourceforge.net/projects/interrogate/. Last visited 24-06-2014
58. http://pdos.csail.mit.edu/6.828/2009/readings/i386/s05_02.htm. Last visited 24-06-2014
59. H. Riebler et al. *FPGA-accelerated Key Search for Cold-Boot Attacks against AES*. In Int. Conf. on Field-Programmable Technology (FPT), 2013
60. RSA Laboratories. *PKCS #1 v2.2: RSA Cryptography Standard*. 2012
61. http://en.wikipedia.org/wiki/Serpent_(cipher) . Last visited 24-06-2014
62. http://en.wikipedia.org/wiki/Twofish. Last visited 24-06-2014
63. http://en.wikipedia.org/wiki/Substitution_box. Last visited 24-06-2014
64. R. Carbone et al. *An in-depth analysis of the cold boot attack*. 2011
65. http://en.wikipedia.org/wiki/DMA_attack. Last visited 24-06-2014
66. M. Gruhn et al. *On the Practicability of Cold Boot Attacks*. In 8th Int. Conf. on Availability, Reliability and Security (ARES), 2013
67. http://en.wikipedia.org/wiki/Cloud_storage. Last visited 24-06-2014
68. T. Müller et al. *FROST: Forensic Recovery of Scrambled Telephones*. In Proceedings of the 11th int. conf. on Applied Cryptography and Network Security, 2013
69. http://en.wikipedia.org/wiki/Dm-crypt. Last visited 24-06-2014
70. http://en.wikipedia.org/wiki/Lock_screen. Last visited 24-06-2014
71. http://en.wikipedia.org/wiki/Smudge_attack. Last visited 24-06-2014
72. A. Skillen et al. *Deadbolt: Locking Down Android Disk Encryption*. In Proceedings of the 3rd ACM workshop on Security and privacy in smartphones & mobile devices, 2013

73. P. McGregor. *Braving the Cold: New Methods for Preventing Cold Boot Attacks on Encryption Keys*. Black Hat US, 2008
74. T. Müller et al. *TRESOR Runs Encryption Securely Outside RAM*. In Proceedings of the 20th USENIX conference on Security, 2011
75. http://en.wikipedia.org/wiki/X86_debug_register. Last visited 24-06-2014
76. http://en.wikipedia.org/wiki/AES_instruction_set. Last visited 24-06-2014
77. http://en.wikipedia.org/wiki/Streaming_SIMD_Extensions. Last visited 24-06-2014
78. T. Müller et al. *TreVisor: OS-Independent Software-Based Full Disk Encryption Secure against Main Memory Attacks*. In Proceedings of the $10^{th}$ int. conf. on Applied Cryptography and Network Security, 2012
79. http://www.bitvisor.org/. Last visited 24-06-2014
80. J. Götzfried et al. *ARMORED: CPU-bound Encryption for Android-driven ARM Devices*. In $8^{th}$ int. conf. on Availability, Reliability and Security (ARES), 2013
81. P. Simmons. *Security Through Amnesia: A Software-Based Solution to the Cold Boot Attack on Disk Encryption*. In Proceedings of the 27th Annual Computer Security Applications Conference, 2011
82. http://sourceforge.net/projects/loop-aes/. Last visited 24-06-2014
83. http://en.wikipedia.org/wiki/Model-specific_register. Last visited 24-06-2014
84. E. Blass et al. *TRESOR-HUNT: Attacking CPU-Bound Encryption*. In Proceedings of the 28th Annual Computer Security Applications Conference, 2012
85. http://en.wikipedia.org/wiki/Interrupt_descriptor_table. Last visited 24-06-2014
86. http://en.wikipedia.org/wiki/IOMMU. Last visited 24-06-2014
87. B. Garmany et al. *PRIME: Private RSA Infrastructure for Memory-less Encryption*. In Proceedings of the 29th Annual Computer Security Applications Conference, 2013
88. http://en.wikipedia.org/wiki/Montgomery_reduction. Last visited 24-06-2014
89. L. Guan et al. *Copker: Computing with Private Keys without RAM*. In Network and Distributed System Security Symposium (NDSS), 2014
90. http://en.wikipedia.org/wiki/Chinese_remainder_theorem. Last visited 24-06-2014
91. R. Canetti et al. *Exposure-Resilient Functions and All-Or-Nothing Transforms*. In Lecture Notes in Computer Science Volume 1807, 2000
92. J. Alawatugoda et al. *Continuous After-the-fact Leakage-Resilient Key*. In Proceedings of the 19th Australasian Conference on Information Security and Privacy, 2014
93. A. Akavia et al. *Simultaneous Hardcore Bits and Cryptography against Memory Attacks*. In Lecture Notes in Computer Science Volume 5444, 2009
94. http://en.wikipedia.org/wiki/One-time_pad. Last visited 24-06-2014

95. M. Henson et al. *Memory Encryption: A Survey of Existing Techniques*. In ACM Computing Surveys (CSUR), Volume 46 Issue 4, 2014
96. G. Suh et al. *AEGIS: A Single-Chip Secure Processor*. In Information Security Tech. Report, Volume 10, Issue 2, 2005
97. http://privatecore.com/vcage/. Last visited 24-06-2014
98. http://en.wikipedia.org/wiki/Hardware-based_full_disk_encryption. Last visited 24-06-2014
99. T. Müller et al. *Self-Encrypting Disks pose Self-Decrypting Risks*. In 29th Chaos Communication Congress, 2012